\documentclass[desactivate]{aa}  
\usepackage{natbib}
\bibpunct{(}{)}{;}{a}{}{,}

\usepackage{graphicx}
\usepackage{revsymb}	
\usepackage{amsmath}
\usepackage{txfonts}
\usepackage{amssymb}
\usepackage{geometry} 
\usepackage[parfill]{parskip} 
\usepackage{slantsc}
\usepackage{array}
\usepackage{amsfonts}

\usepackage{epstopdf}	
\usepackage{epsfig}	

\usepackage{url}

\usepackage[colorlinks=true,linkcolor=black, urlcolor=black, citecolor=black]{hyperref}

\usepackage{color}

\usepackage{xfrac}

\usepackage{sidecap}
\usepackage[font=small, labelfont=bf]{caption}
\usepackage{subcaption}

\usepackage{multirow}
\usepackage{booktabs}
\usepackage{threeparttable}

\begin{document}
   \title{Minimising magnetic activity effects in PLATO observations: insights from the Sun-as-a-star}
\author{J. B\'{e}trisey\inst{1} \and A. -M. Broomhall\inst{2} \and S.~N. Breton\inst{3} \and E.~Panetier \inst{4} \and R.~A. Garc{\'\i}a\inst{4} \and H. Davenport\inst{2,5}\and O. Kochukhov\inst{1}}
\institute{Department of Physics and Astronomy, Uppsala University, Box 516, SE-751 20 Uppsala, Sweden\\email: 	\texttt{jerome.betrisey@physics.uu.se}
\and Centre for Fusion, Space and Astrophysics, Department of Physics, University of Warwick, Coventry CV4 7AL, UK
\and  INAF – Osservatorio Astrofisico di Catania, Via S. Sofia 78, 95123 Catania, Italy
\and Université Paris Cité, Université Paris-Saclay, CEA, CNRS, AIM, 91191 Gif-sur-Yvette, France
\and Université Paris-Saclay, Université Paris Cité, CEA, CNRS, AIM, 91191 Gif-sur-Yvette, France
\and Blackett Laboratory, Imperial College London, London SW7 2AZ, United Kingdom}
\date{\today}

\abstract
{Asteroseismic modelling will be a key component of upcoming missions like PLATO, CubeSpec, and \textit{Roman}. Magnetic activity effects in solar‑type stars is often assumed negligible, with its influence absorbed into surface corrections, yet recent studies show that it can substantially bias seismic inferences, particularly age estimates, regardless of modelling strategy or surface treatment.}
{We quantified how magnetic activity effects in the Sun-as-a-star are smoothed by temporal averaging by analysing 182.5-, 365-, 730-, and 1460-day time series from the BiSON network and the GOLF instrument.}
{For each baseline, we analysed overlapping segments to track the apparent temporal evolution of the seismic age and its correlation with the 10.7\,cm radio flux. We then estimated the associated systematic uncertainty using two metrics from the literature. We finally compared results across baselines to evaluate how the observing window shapes activity‑induced biases.}
{Across all 28 tested configurations, solar‑cycle signatures persist even in 1460‑day windows. Under these conditions, activity‑driven systematics reach $\sim$1--2.5\% for fits to individual frequencies and $\sim$2.5--4.5\% when using frequency separation ratios. With a two‑year baseline, the corresponding biases reach $\sim$2--3\% and $\sim$3.5--5.5\%. The suppression of magnetic activity effects with increasing baseline is non‑monotonic: one‑ and four‑year windows reduce biases far more effectively than shorter baselines in most cases, whereas 730‑day windows provide only limited improvement over 365‑day ones. Improvements arise from enhanced frequency determination (dominant at 365 days) and from increasingly efficient temporal averaging of the activity cycle (dominant at 1460 days). In contrast, 730‑day is an intermediate regime: frequency accuracy has already plateaued and the observing window remains too short to smooth out cycle‑related variability. On average, we find that magnetic activity effects decrease by 24\%, 12\%, 30\%, and 38\% when transitioning from 182.5 to 365 days, 365 to 730 days, 730 to 1460 days, and 365 to 1460 days for frequency‑based fits; the corresponding improvements for ratio‑based fits are 13\%, 14\%, 20\%, and 31\%.}
{These results indicate that a continuous single-field four‑year PLATO observing programme would provide the most effective suppression of magnetic‑activity biases for solar analogues, whereas a 2+2‑year strategy (in two distinct fields) is significantly more sensitive to magnetic effects, with limited gains between 365‑ and 730‑day windows.}


\maketitle

\defcitealias{Betrisey2023_AMS_surf}{JB23}
\defcitealias{Betrisey2024_MA_Sun}{JB24}
\defcitealias{Betrisey2025_MA_Inv}{JB25b}
\defcitealias{Betrisey2025_MA_Damping}{JB25a}

\section{Introduction}
Turbulent convection in the upper layers of solar-like stars stochastically drives a broad spectrum of global oscillations. Analysing these mode patterns opens a direct window into stellar interiors and sets tight constraints on the fundamental stellar parameters, such as mass, radius, and age. These constraints are pivotal for robust models of stellar and planetary system evolution and for galactic archaeology \citep[see e.g. the reviews by][]{Chaplin&Miglio2013,Garcia&Ballot2019,Aerts2021}. Asteroseismology has been revolutionised by space-based photometry missions such as Convection, Rotation and planetary Transits \citep[CoRoT;][]{Baglin2009}, \textit{Kepler} \citep{Borucki2010}, K2 \citep{Howell2014}, and Transiting Exoplanet Survey Satellite \citep[TESS;][]{Ricker2015}. Their legacy has set the stage for a new generation of space telescopes, including PLAnetary Transits and Oscillations of stars \citep[PLATO;][]{Rauer2025} and CubeSpec \citep{Bowman2022}, for which asteroseismic inferences will be a central pillar. Achieving the ambitious accuracy targeted by PLATO (15\% in mass, 1-2\% in radius, and 10\% in age for a distant sun) highlights the need of addressing persistent mismatches between observed pulsation spectra and theoretical predictions, which can significantly bias seismic inferences at that precision level.

Current limitations in asteroseismic modelling arise primarily from three fronts: 1) uncertainties in the microphysics and macrophysics employed in stellar models \citep[e.g.][]{Buldgen2019f,Farnir2020,Betrisey2022}, 2) inadequate treatment of near‑surface layers \citep[e.g.][]{Ball&Gizon2017,Nsamba2018,Jorgensen2020,Jorgensen2021,Cunha2021,Betrisey2023_AMS_surf,Betrisey2026_catalog}, and 3) the influence of stellar magnetic activity \citep[e.g.][]{Garcia2010,Broomhall2011,Santos2018,PerezHernandez2019,Santos2019_sig,Santos2019_rot,Howe2020,
Thomas2021,Santos2021,Santos2023,Betrisey2024_MA_Sun,Betrisey2025_MA_Inv,Betrisey2025_MA_Damping}. The latter produces systematic shifts in mode frequencies. First identified in solar observations of low‑degree modes by \citet{Woodard&Noyes1985}, this phenomenon has since been firmly established through numerous studies \citep[for a review, see e.g.][]{Broomhall&Nakariakov2015}. Beyond the dominant main cycle, quasi‑biennial oscillation shifts have also been detected both in solar data \citep[e.g.][]{Broomhall2009_lack,Fletcher2010,Kolotkov2015,Mehta2022,Jain2023} and in other solar-like oscillators \citep[e.g.][]{Santos2019_sig}. Although the physical origin of frequency shifts due to magnetic activity is not yet settled, two explanations dominate the literature: temporal variations of magnetic fields in the outer stellar layers \citep[e.g.][]{Howe2002,Baldner2009,Garcia2024}, and structural changes within the acoustic cavity itself \citep[e.g.][]{Woodard&Noyes1985,Fossat1987,Libbrecht&Woodard1990,Kuhn1998,Dziembowski&Goode2005,Basu2012}.

The impact of magnetic activity is not limited to individual mode frequencies. Variations over the activity cycle have also been observed in global seismic parameters, such as the frequency of maximum oscillation power \citep{Howe2020} and the large frequency separation \citep[e.g.][]{Broomhall2011}. As a consequence, asteroseismic scaling relations \citep[see e.g.][for a review]{Hekker2020} are directly biased by magnetic activity effects. These biases also propagate into inferred stellar properties, including mass, radius, age, mean density, acoustic radius, and the initial helium mass fraction. This conclusion holds irrespective of the modelling framework, whether forward modelling approaches \citep[e.g.][hereafter JB24]{Creevey2011,PerezHernandez2019,Thomas2021,Betrisey2024_MA_Sun}, inversion‑based techniques \citep[][hereafter JB25b]{Betrisey2025_MA_Inv}, or surface‑independent methods \citep[][hereafter JB25a]{Betrisey2025_MA_Damping} are employed. In particular, seismic age determinations are the most sensitive to magnetic activity, with substantial biases at the precision level targeted by PLATO.

Because magnetic activity biases seismic inferences regardless of the modelling strategy adopted, the present work expands our recent investigations of this problem \citepalias[using 365-day observing baselines;][]{Betrisey2024_MA_Sun,Betrisey2025_MA_Inv,Betrisey2025_MA_Damping} by examining how these activity-induced biases evolve when the observational time span is varied. As pointed out in our earlier studies, Sun-as-a-star data provide an ideal benchmark for this purpose: the Sun is the only star for which multi‑decade, cycle‑spanning observations of acoustic oscillations exist, and the accuracy requirements of PLATO are defined specifically for a solar analogue. In Sect.~\ref{sec:summary_literature}, we summarise our previous investigations with 365-day data and introduce the observational datasets used for this study. In Sect.~\ref{sec:imprint_MA_cycle}, we investigate the correlation between the solar activity cycle and the seismic age obtained from both surface‑dependent and surface‑independent methods. Section~\ref{sec:averaging_observing_baselines} then quantifies to what extent extended observing windows attenuate magnetic‑activity systematics. Our conclusions are summarised in Sect.~\ref{sec:conclusions}.

\section{Previous investigations and observational datasets}
\label{sec:summary_literature}
The present study extends our previous investigations based on 365-day segments of the Sun as a star \citepalias{Betrisey2024_MA_Sun,Betrisey2025_MA_Inv,Betrisey2025_MA_Damping}. Following the approach adopted in our earlier works, our analysis draws on two independent sets of Sun-as-a-star Doppler velocity observations: the ground-based BiSON network \citep[Birmingham Solar Oscillations Network;][]{Davies2014,Hale2016} and the space-based GOLF instrument \citep[Global Oscillations at Low Frequencies;][]{Gabriel1995}. The use of these complementary datasets, spanning more than three decades of continuous observations of the Sun, ensures that our conclusions are robust and not instrument-dependent. To probe the temporal evolution of global solar properties, the BiSON and GOLF time series are divided into shorter, partially overlapping segments. Within each segment, we extract acoustic mode frequencies and infer fundamental stellar parameters, including mass, radius, and age.

In \citetalias{Betrisey2024_MA_Sun}, we investigated the impact of magnetic activity using surface-dependent modelling, exploring several semi-empirical prescriptions \citep{Kjeldsen2008,Ball&Gizon2014,Sonoi2015}. Although part of the activity-induced signal can be absorbed by the surface-effect parametrisation, we identified a clear residual imprint on the inferred stellar properties. In particular, the seismic age exhibits significant variations over the solar cycle, in contrast with previous expectations \citep[e.g.][]{Howe2017}. This behaviour is consistently observed in both BiSON and GOLF datasets and is independent of the adopted surface correction. The inferred solar age varies by up to 6.5\% between activity extrema, corresponding to age differences of up to 300 Myr. Moreover, the effect is more pronounced during the more active Cycle 23. Given that the Sun is not particularly active \citep[e.g.][]{Santos2023}, these results highlight magnetic activity as a potentially significant source of bias for the characterisation of more active stars, such as those that will be observed by PLATO. It also suggests that active \textit{Kepler} targets may require reanalysis accounting for such effects. Additionally, we explored mitigation strategies, finding that lower activity levels (e.g. Cycle 24) naturally reduce biases. The inclusion of low-order modes ($n=12-15$), which are less sensitive to activity effects, provides moderate improvements. However, detecting these modes with PLATO will be quite challenging due to the photometric background.

In \citetalias{Betrisey2025_MA_Inv}, we extended the analysis of \citetalias{Betrisey2024_MA_Sun} using inverse techniques, specifically mean density and acoustic radius inversions \citep[see e.g.][for reference articles and a recent review]{Reese2012,Buldgen2015a,Buldgen2022c}, obtaining consistent conclusions. Since the two main types of seismic inferences are affected by magnetic activity effects, the development of mitigation strategies becomes essential. We identified four promising approaches: 1) applying the damping method for surface effects based on surface-independent constraints; 2) increasing the observing window to average out cycle-related variations; 3) pre-correcting oscillation frequencies for activity-induced shifts; and 4) developing surface-effect-like prescriptions tailored to absorb activity-related perturbations in the frequency-ratio space. The first option was investigated in \citetalias{Betrisey2025_MA_Damping}, while the present study focuses on the second; the remaining approaches will be explored in future works.

\citetalias{Betrisey2025_MA_Damping} showed that frequency separation ratios do not suppress magnetic activity biases in asteroseismic inferences. This stems from the fact that activity-induced perturbations are asymmetric with respect to the harmonic degree $l$. Consequently, the activity signal is also present in the frequency separation ratios and then propagates to the inferred stellar parameters during the fitting process. This behaviour contrasts with that of surface effects, which are largely symmetric with respect to $l$. At a given frequency, the associated frequency shifts are approximately constant across modes of different harmonic degree (e.g. radial, dipolar, and quadrupolar modes). By constructing suitable combinations of frequencies with different $l$, it is therefore possible to efficiently cancel out surface effects \citep[see e.g.][]{Roxburgh&Vorontsov2003,Oti2005}. These combinations, known as frequency separation ratios, can thus be used as so-called surface-independent constraints in seismic modelling. Methods relying on such constraints are referred to as surface-independent modelling, whereas approaches based on individual frequencies, which remain sensitive to surface effects, are termed surface-dependent modelling.

\citetalias{Betrisey2024_MA_Sun} showed that, when using surface-dependent modelling, the activity-induced bias predominantly affects the inferred stellar age. This behaviour arises from the inclusion of additional free parameters in the surface-effect parametrisation, which partially absorb the activity signal. As a result, the impact on most other stellar parameters is significantly reduced. In contrast, when surface-independent constraints are used, no such absorption mechanism is available. The activity-induced bias therefore propagates more directly to the inferred parameters. While most stellar properties (e.g. mass and radius) remain only weakly affected, the seismic age exhibits the strongest correlation with the activity cycle. The resulting systematic uncertainty on the inferred age is slightly larger in the surface-independent case ($\sim 4.7\%$) than in surface-dependent approaches ($\sim 3\%$). Nevertheless, surface-dependent modelling remains significantly more affected by inaccuracies in the treatment of surface effects \citep[e.g.][]{Nsamba2018,Jorgensen2020,Jorgensen2021,Cunha2021,Betrisey2023_AMS_surf,Betrisey2026_catalog}. For this reason, we strongly advocate the use of surface-independent methods, as their robustness against surface effects outweighs their increased sensitivity to magnetic activity.

To investigate the dependence of magnetic-activity effects on the temporal coverage, both datasets were segmented into four observing baselines of 182.5, 365, 730, and 1460\,days. These segments were constructed with overlapping windows, using offsets of 91.25, 91.25, 182.5, and 365\,days, respectively. The procedures used to extract oscillation frequencies from the BiSON and GOLF timeseries are described in detail by \citet{Garcia2005}, \citet{Breton2022_astro}, and \citetalias{Betrisey2025_MA_Damping}. The BiSON spectra covers modes up to approximately 3600 $\mu$Hz, while GOLF extends slightly further, reaching $\sim3800$ $\mu$Hz \citepalias[see][]{Betrisey2025_MA_Damping}. This difference has a negligible impact on our analysis. The slightly higher data quality of BiSON compensates for the absence of a small number of highest-order modes, yielding comparable diagnostic power between the two datasets.  BiSON serves as the reference dataset and GOLF data are used to confirm that the identified trends are instrument-independent. In this context, analysing the 365-, 730-, and 1460-day GOLF baselines is sufficient; including 182.5-day GOLF segments would add substantial computational cost with little additional scientific value, and is therefore not pursued here. But both BiSON and GOLF results are used for quantitative results whenever both are available for a given baseline.

\section{Imprint of the magnetic activity cycle}
\label{sec:imprint_MA_cycle}
\begin{table}[t!]
\centering
\caption{Spearman correlation coefficient between seismic age and the 10.7~cm radio emission flux, a proxy of the solar cycle.}
\resizebox{0.99\linewidth}{!}{
\begin{tabular}{lcccc}
\hline \hline
Observing time & \multicolumn{2}{c}{BiSON} & \multicolumn{2}{c}{GOLF} \\
 \cmidrule[0.4pt](lr){2-3} \cmidrule[0.4pt](l){4-5}
 & $n \geq 12$ & $n \geq 16$ & $n \geq 12$ & $n \geq 16$  \\
\hline 
\textit{Fit frequencies} & & & & \\
182.5 days & $0.33\pm 0.17$ & $0.43\pm 0.07$ & - & - \\ 
365 days & $0.57\pm 0.09$ & $0.64\pm 0.07$ & $0.46\pm 0.11$ & $0.56\pm 0.14$ \\ 
730 days & $0.71\pm 0.20$ & $0.79\pm 0.10$ & $0.52\pm 0.17$ & $0.62\pm 0.16$ \\ 
1460 days & $0.72\pm 0.21$ & $0.78\pm 0.16$ & $0.69\pm 0.23$ & $0.67\pm 0.10$ \\ 
\hline 
\textit{Fit ratios} & & & & \\
182.5 days & $0.45\pm 0.02$ & $0.37\pm 0.06$ & - & - \\ 
365 days & $0.62\pm 0.07$ & $0.61\pm 0.11$ & $0.56\pm 0.09$ & $0.52\pm 0.08$ \\ 
730 days & $0.67\pm 0.13$ & $0.67\pm 0.03$ & $0.59\pm 0.08$ & $0.61\pm 0.04$ \\ 
1460 days & $0.70\pm 0.20$ & $0.80\pm 0.14$ & $0.36\pm 0.50$ & $0.67\pm 0.10$ \\ 
\hline 
\end{tabular}}
{\par\small\justify\textbf{Notes.}  The uncertainties were computed following the methodology introduced by \citetalias{Betrisey2024_MA_Sun}. Specifically, they quantify the statistical stability of the inferred Spearman correlation coefficient. \par}
\label{tab:correlations}
\end{table}

We estimated the seismic age of the Sun-as-a-star using two complementary inference approaches. The first relies on a direct fit of the individual mode frequencies together with spectroscopic constraints and is therefore sensitive to the surface effects \citep[see][hereafter JB23]{Betrisey2023_AMS_surf}. Specifically, this corresponds to the initial modelling step of the Forward and Inverse COmbination pipeline \citep[FICO;][]{Betrisey2022,Betrisey2023_AMS_surf,Betrisey2024_AMS_quality,Betrisey2026_catalog}. The second method suppresses surface-effect systematics \citepalias[but not magnetic activity effects as shown in][]{Betrisey2025_MA_Damping} by fitting ratios of frequencies and incorporating the inverted mean density in the constraints. This corresponds to the complete FICO procedure \citepalias[see][and our previous works]{Betrisey2023_AMS_surf}. As illustrative examples, this procedure has been applied in several studies to improve the characterisation of solar-like oscillators \citep[see e.g.][]{Betrisey2022,Betrisey2023_AMS_surf,Betrisey2026_catalog,
Buldgen2025_lithium,Pezzotti2026,Garcia2026}. Following our earlier studies, we also explored two mode sets: one including low-order modes ($n \geq 12$), and one restricted to higher radial orders only ($n \geq 16$). Although the detection of low-order modes may be more challenging for missions such as PLATO \citepalias[see discussion in][]{Betrisey2024_MA_Sun}, these modes are known to be less affected by magnetic activity \citep[e.g.][]{Broomhall&Nakariakov2015} and thus help reduce magnetic-activity biases \citepalias{Betrisey2024_MA_Sun,Betrisey2025_MA_Inv,Betrisey2025_MA_Damping}. Combining the various instruments, observing baselines, inference methods, and mode selections yields a total of 28 distinct seismic-age datasets studied in this work.

\begin{figure}[t!]
\centering
\begin{subfigure}[b]{.41\textwidth}
  \includegraphics[width=.99\textwidth]{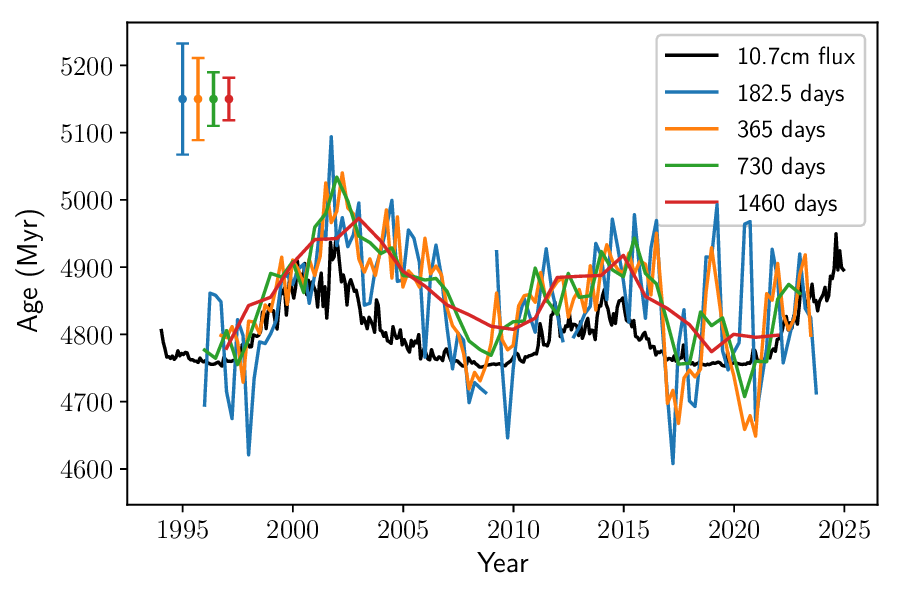}  
\end{subfigure}
\begin{subfigure}[b]{.41\textwidth}
  \includegraphics[width=.99\textwidth]{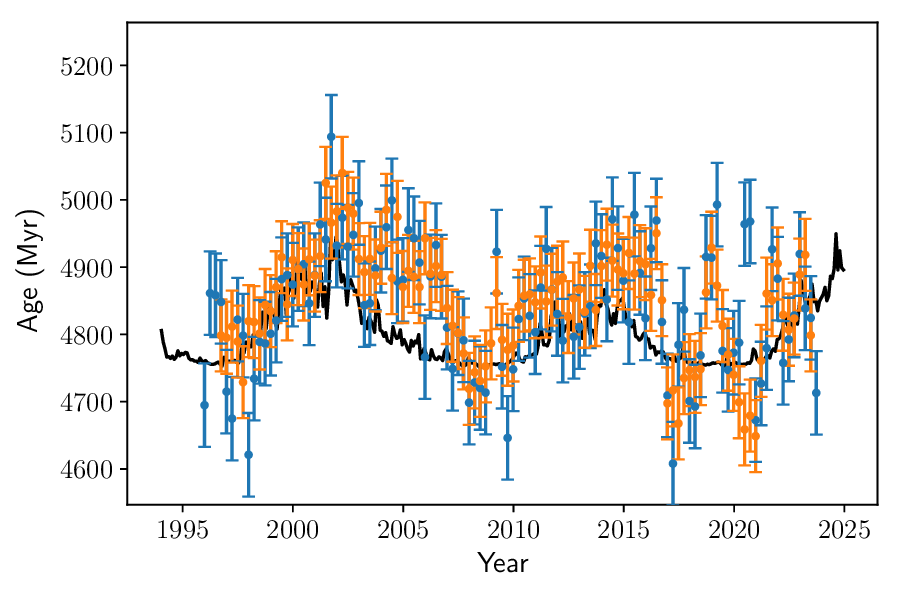}  
\end{subfigure}
\begin{subfigure}[b]{.41\textwidth}
  \includegraphics[width=.99\textwidth]{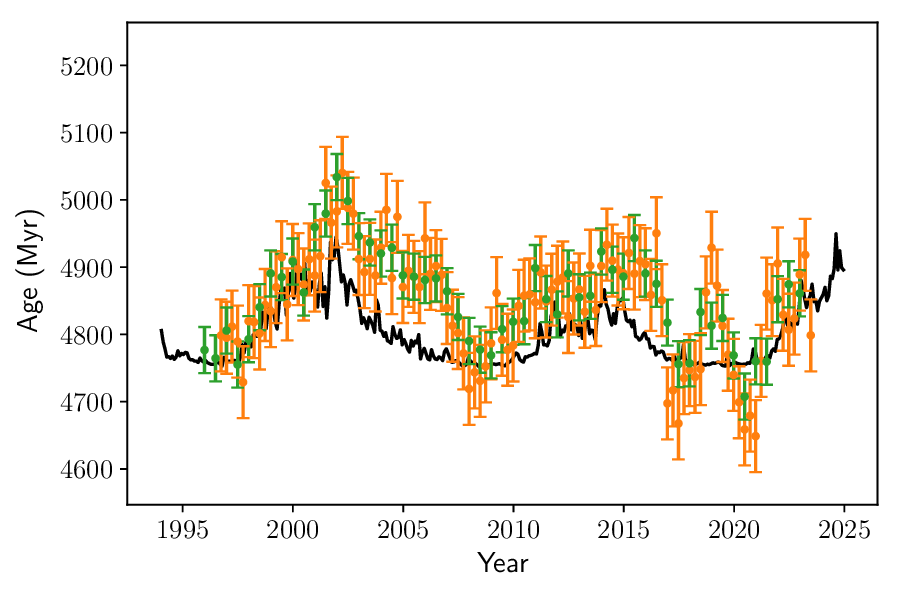}  
\end{subfigure}
\begin{subfigure}[b]{.41\textwidth}
  \includegraphics[width=.99\textwidth]{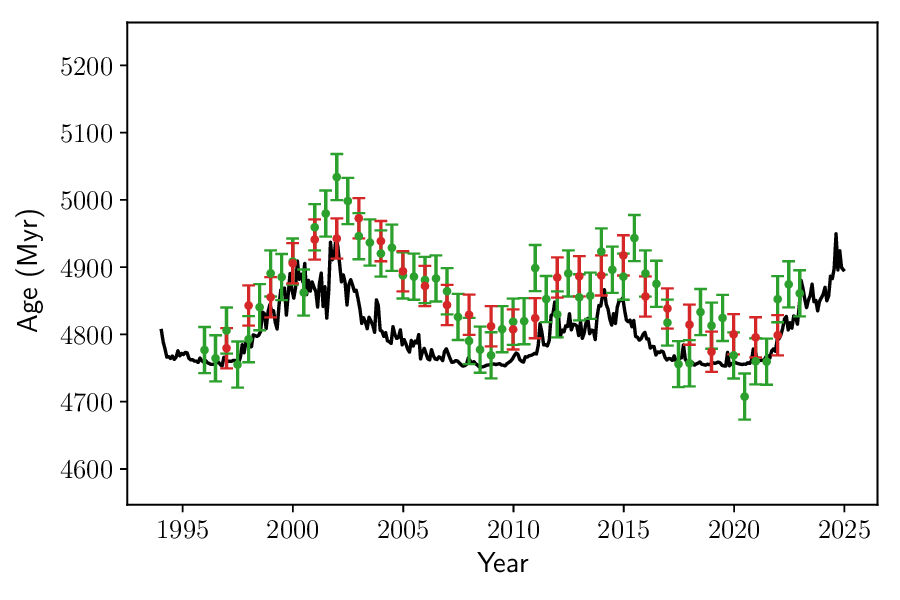}  
\end{subfigure}
\caption{Temporal evolution of seismic age for different observing times with the BiSON network and using modes with $n\geq 16$. The first panel includes the four observing windows without errorbar for readability purposes. The second panel compares 182.5 days with 365 days results, the third panel 365 days with 730 days, and the fourth panel 730 days with 1460 days. The black line represents the (rescaled) 10.7 cm radio emission flux, a standard proxy of the solar cycle.}
\label{fig:bison_errorbars}
\end{figure}

Following the methodology of \citetalias{Betrisey2024_MA_Sun} to properly account for correlations induced by overlapping data segments, we computed the Spearman correlation coefficient \citep{Spearman1904} between the seismic age and the 10.7~cm radio flux\footnote{\url{https://www.spaceweather.gc.ca/}}, a standard proxy of solar magnetic activity. The results are summarised in Table~\ref{tab:correlations}. Illustrative plots with BiSON datasets are presented in Fig.~\ref{fig:bison_errorbars}. All datasets exhibit a clear imprint of the solar cycle, including those based on the longest, 1460-day observing windows. We also note that, in most cases, correlations strengthen as the time series become longer, reflecting the impact of improved frequency determination.

For frequency-based fits, adding the four low-order radial modes reduces the correlation with activity, in line with \citetalias{Betrisey2024_MA_Sun}, and this trend holds in nearly all cases, except for the 1460‑day GOLF data, where the correlation coefficients remain essentially unchanged. For the surface-independent fits using frequency separation ratios, no similarly simple pattern emerges regarding the inclusion or exclusion of the lowest-order modes. This is expected: as shown in \citetalias{Betrisey2025_MA_Damping} for the 365-day case, surface-independent fits redistribute magnetic-activity perturbations differently because the surface term is handled via the usage of surface-independent constraints rather than through additional free parameters. When directly fitting frequencies, the free parameters defining the surface-effect prescription absorb a substantial portion of the magnetic signal, leaving the seismic age as the only main stellar parameter exhibiting cycle-related variations. In contrast, when fitting ratios, all optimised parameters respond to magnetic activity in a non-trivial, highly coupled fashion. In this regime, the inferred age, mass, and initial helium abundance influence and contaminate one another, with the seismic age still showing the strongest response but with the other parameters absorbing variable fractions of the magnetic signal depending on the chosen constraint set. As a consequence, simple, monotonic patterns like those found for the fits of individual frequencies do not occur for the ratio-based inferences.

\section{Evolution of systematic uncertainty due to magnetic activity}
\label{sec:evolution_MA_uncertainty}
This section examines how magnetic-activity-induced biases evolve as a function of the observing baseline. To quantify the associated systematic uncertainty, we adopted two metrics commonly used in the literature. The first (labelled as `standard systematic') follows the definition introduced by \citetalias{Betrisey2025_MA_Inv} and \citetalias{Betrisey2025_MA_Damping}, which estimates the bias based on the standard deviation of the seismic-age dataset. The second (labelled as `extrema systematic') follows \citetalias{Betrisey2024_MA_Sun} and uses (half of) the range spanned by the dataset extrema. While the latter provides a conservative, worst-case estimate, it is highly sensitive to outliers. The former was specifically designed to deliver a comparable uncertainty estimate while mitigating the influence of outliers, and to remain easily applicable to space-based observations.

\begin{figure}[t!]
\centering
\begin{subfigure}[b]{.41\textwidth}
  \includegraphics[width=.99\textwidth]{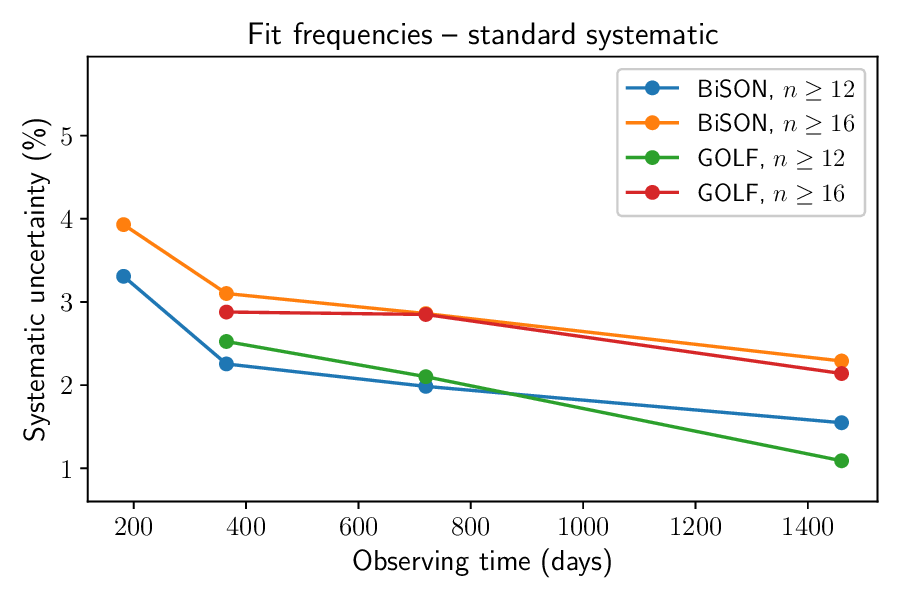}  
\end{subfigure}
\begin{subfigure}[b]{.41\textwidth}
  \includegraphics[width=.99\textwidth]{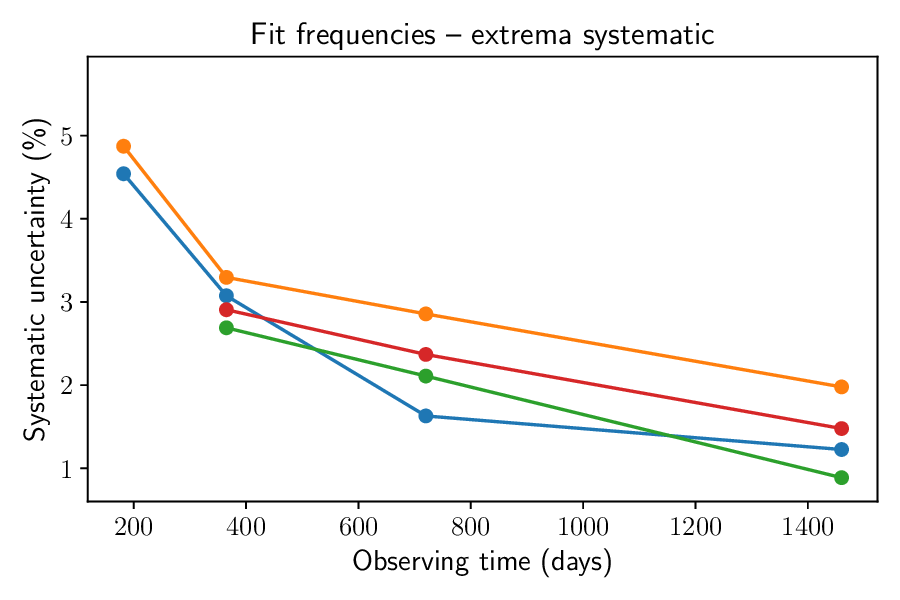}  
\end{subfigure}
\begin{subfigure}[b]{.41\textwidth}
  \includegraphics[width=.99\textwidth]{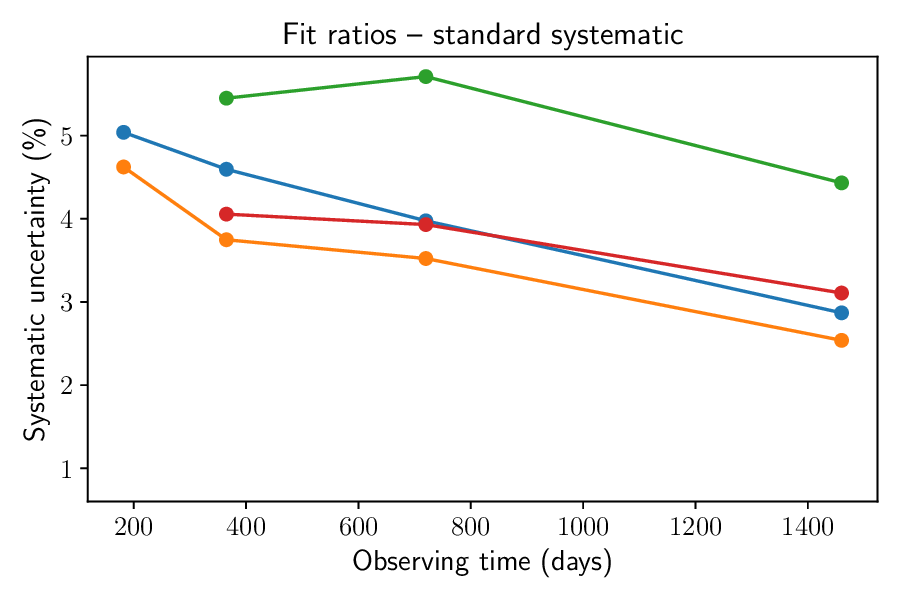}  
\end{subfigure}
\begin{subfigure}[b]{.41\textwidth}
  \includegraphics[width=.99\textwidth]{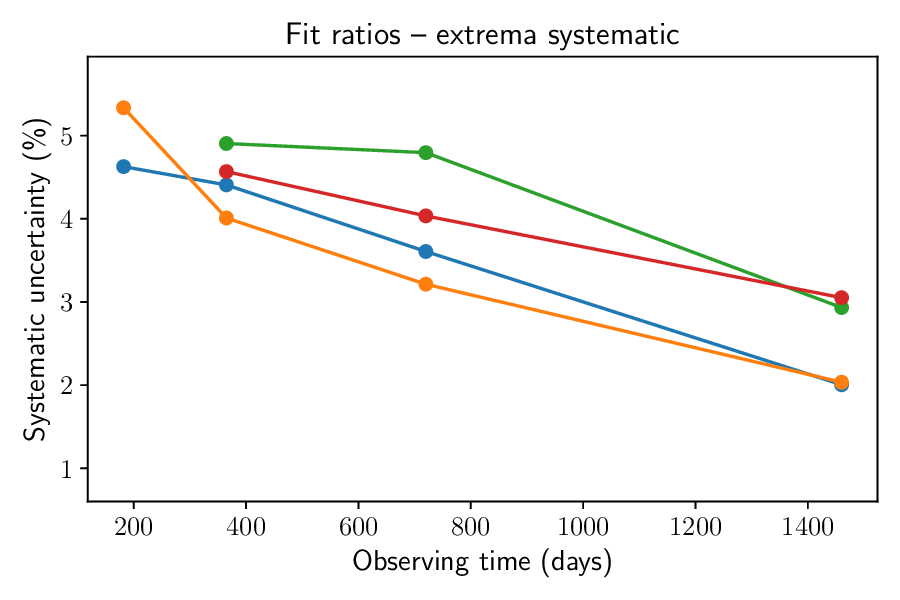}  
\end{subfigure}
\caption{Evolution of systematic uncertainty due to magnetic activity for solar cycle 23 and for the different datasets (blue to red) in our study. The seismic age was estimated using two methodologies: surface-dependent (labelled as `fit frequencies') and surface-independent (labelled as `fit ratios') methods \citepalias[see][]{Betrisey2025_MA_Damping}. The systematic uncertainty was estimated using two methods: the definition from \citetalias{Betrisey2025_MA_Inv} and \citetalias{Betrisey2025_MA_Damping} based on the standard deviation of the dataset (labelled as `standard systematic'), and the definition from \citetalias{Betrisey2024_MA_Sun} based on the dataset extrema (labelled as `extrema systematic').}
\label{fig:MA_systematic_cycle23}
\end{figure}

Figure~\ref{fig:MA_systematic_cycle23} shows the evolution of both uncertainty metrics as a function of the observing baseline for all tested modelling configurations. Results for Cycle 24 are provided in Appendix~\ref{app:evolution_MA_uncertainty}. For both activity cycles, the systematic uncertainty decreases as the observing baseline increases. This behaviour indicates that magnetic activity biases are progressively averaged out when longer time series are used. While this trend is expected on general grounds, it is nevertheless reassuring to confirm it explicitly in the context of preparing future space-based photometric missions such as PLATO. We note, however, three departures from this trend (see e.g. the green curve in the bottom-left panel of Fig.~\ref{fig:MA_systematic_cycle23} and the negative values in Table~\ref{tab:improvement_factor_full} for the complete list). In these cases, the 730-day baseline yields a slightly larger systematic uncertainty than the 365-day baseline.

Although activity-induced biases can be mitigated with longer observing baselines, a discernible imprint of the solar cycle remains even when using the longest 1460-day time series. For this baseline, the residual systematic uncertainty induced by magnetic activity is at the level of $\sim$1--2.5\% when fitting individual oscillation frequencies, and $\sim$2.5--4.5\% when using frequency separation ratios. These values are well below the 10\% age-accuracy requirement set by PLATO for Sun-like stars. However, they remain non-negligible in the context of high-precision asteroseismology.

A more global underperformance of the 730-day datasets can also be identified. Beyond the three outliers mentioned above, the relative reduction in systematic uncertainty is generally smaller when increasing the baseline from 365 to 730 days than for other transitions, in particular from 182.5 to 365 days and from 730 to 1460 days. This behaviour is systematically quantified in the next section, but it is already visible in Fig.~\ref{fig:MA_systematic_cycle23}. For example, the relative improvement between 730 and 1460 days is typically larger than that between 365 and 730 days. In many of the cases shown in Fig.~\ref{fig:MA_systematic_cycle23}, the trend appears nearly linear across the range spanned by these three baselines, implying a comparatively weaker relative gain at the intermediate duration. As further discussed in the next section, this behaviour reflects the limitations of that observing window. By 730 days, the improvement in frequency accuracy has largely saturated, yet the window remains too short relative to the activity‑cycle timescale to meaningfully average over magnetic variability.

For clarity, we also stress the distinction between accuracy and precision in this context. The 730‑day seismic‑age data naturally achieve higher precision than the 365‑day segments (see upper left panel of Fig.~\ref{fig:bison_errorbars}), although the improvement is smaller than the nominal $\sqrt{2}$ gain in frequency precision expected from a doubling of the observing time. In terms of accuracy, however, both baselines are already long enough that the width of the mode peaks is well resolved and no longer affect the inferred age. The resulting plateau between the
1‑, 2‑, and 4‑year windows arises because, for the mode sets used in our study, the extracted frequencies differ only marginally across these baselines. In the MCMC optimisation, such small shifts lie below the sensitivity of the fit, leading the algorithm to converge to almost identical stellar parameters, with the primary difference being a slightly narrower posterior distribution (and therefore a smaller statistical uncertainty) for longer datasets. This behaviour is reinforced by our use of a consistent set of modes for all observing lengths: if additional lower‑ or higher-order modes could be reliably extracted from the longer segments, their inclusion would likely introduce more significant changes in the inferred parameters. This stands in contrast to the 182.5‑day datasets, for which the limited observing time does lead to a non‑negligible degradation in accuracy.

We also recall that the standard‑deviation‑based metric introduced by \citetalias{Betrisey2025_MA_Inv} and \citetalias{Betrisey2025_MA_Damping} was originally developed using 365‑day datasets. From Figs.~\ref{fig:MA_systematic_cycle23} and \ref{fig:MA_systematic_cycle24}, we find that, across most configurations, this metric yields uncertainty estimates that closely match those obtained from the extrema‑based approach. This agreement indicates that the standard‑deviation‑based metric remains robust when applied to the 730‑day and 1460‑day observing windows, despite the reduced number of data points associated with these longer baselines. A priori, one might have expected the additional variability introduced by the smaller statistics to compromise the applicability of the standard‑deviation‑based metric at these baselines, but our results show that this is not a concern with Sun-as-a-star observations.

\section{Averaging of magnetic activity effects}
\label{sec:averaging_observing_baselines}

\begin{figure*}[t!]
\centering
\includegraphics[scale=0.45]{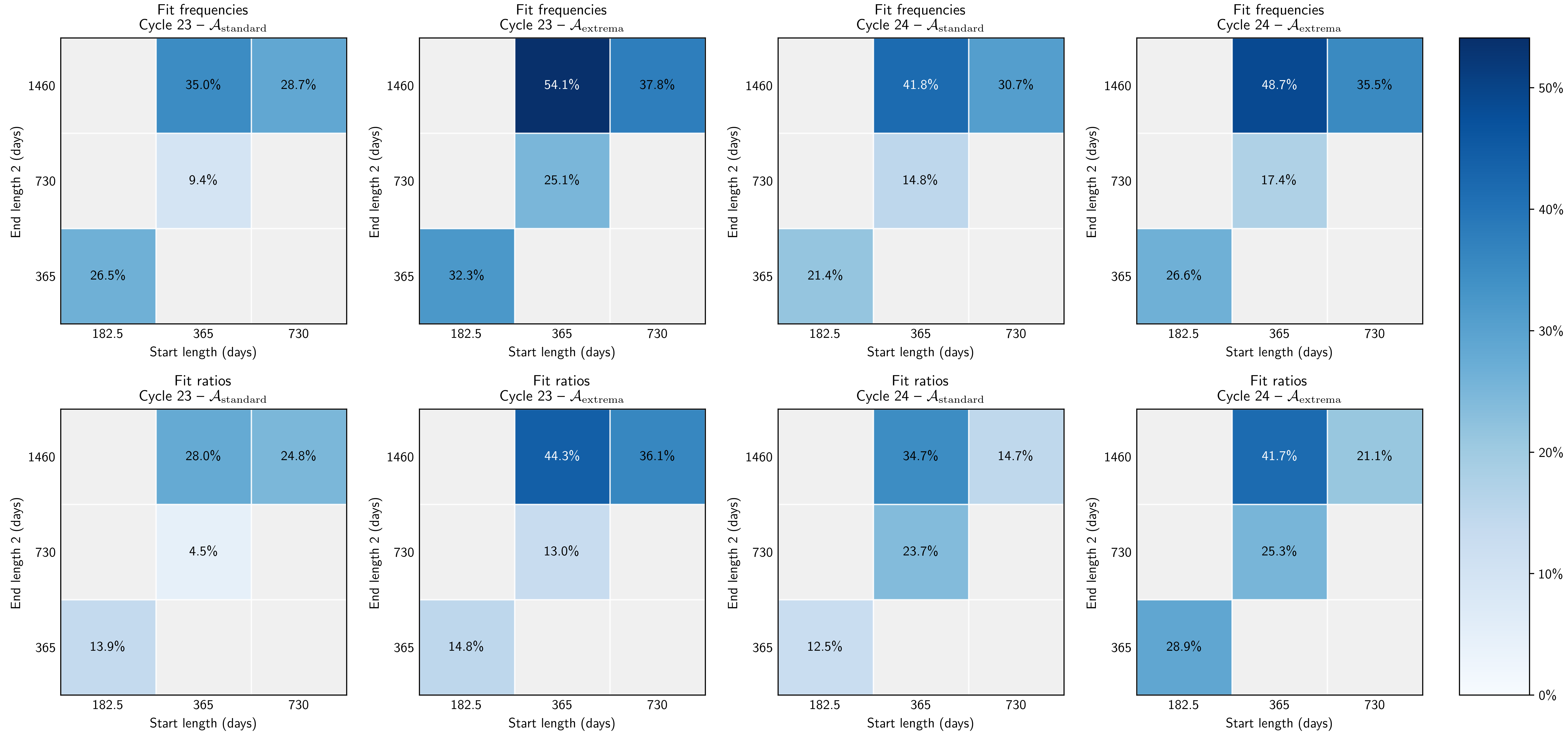} 
\caption{Averaged improvement factor when increasing observing time for Cycles 23 and 24 and for the two metrics that were investigated ($\mathcal{A}_{\mathrm{standard}}$ and $\mathcal{A}_{\mathrm{extrema}}$). Cells show the improvement in percent when transitioning from a shorter (start length) to a longer (end length) time series.}
\label{fig:improvement_factor_heatmap}
\end{figure*}

To characterise how magnetic-activity effects are progressively smoothed out as the observing window increases, we define the following indicator:
\begin{align}
\mathcal{A} = 1 - \frac{\sigma_a}{\sigma_b},
\end{align}
where $\sigma$ denotes the systematic uncertainty associated with magnetic activity, evaluated for a given observing baseline. The subscripts $a$ and $b$ refer to two different baselines, with $a > b$, such that $\sigma_a$ corresponds to the uncertainty measured using a longer time series and $\sigma_b$ to that obtained from a shorter one. By construction, $\mathcal{A}$ therefore quantifies the fractional reduction in magnetic-activity-induced systematics when increasing the duration of the observations from $b$ to $a$. For example, if the uncertainty decreases from $\sigma_b$ for the shorter baseline to $\sigma_a = 0.7\sigma_b$ for the longer one, then $\mathcal{A} = 0.3$, corresponding to a 30\% reduction in the impact of magnetic activity. Conversely, $\mathcal{A}$ can take negative values when $\sigma_a > \sigma_b$, indicating that the longer observing window leads to an increased level of systematic uncertainty. Although rare, this situation was indeed observed in some configurations when transitioning from the 365-day to the 730-day baseline (see Table~\ref{tab:improvement_factor_full}). Detailed numerical results for this improvement factor are provided in Appendix~\ref{app:detailed_table_improvement_factor}, while Fig.~\ref{fig:improvement_factor_heatmap} summarises the corresponding average values for ease of comparison.

We identified three regimes in the behaviour of magnetic-activity mitigation: 1) 182.5 to 365 days, 2) 365 to 730 days, and 3) 365 or 730 to 1460 days. Overall, seismic ages inferred from 365‑day or 1460‑day time series display a markedly stronger suppression of magnetic activity effects compared with shorter baselines. The physical origin of the improvement, however, differs between these two regimes. Both improved frequency determination and the increased temporal smoothing of the activity cycle can reduce the imprint of magnetic activity. In the first regime, where both observing windows remain much shorter than the 11‑year solar cycle, the gain is dominated by improved frequency accuracy. In contrast, the third regime is driven primarily by the smoothing effect, as multi-year windows average over substantial fractions of the cycle. An interesting feature is the underperformance of the 730‑day datasets relative to the 365‑day ones in most of the tested configurations, with noticeably smaller improvement factors across most configurations. This intermediate regime provides little additional benefit: the improvement in frequency accuracy has already plateaued, but a 730‑day window is too short relative to the cycle to significantly smooth out activity variations (see also Appendix~\ref{app:evolution_MA_uncertainty}).

Improvement factors are smaller when using frequency separation ratios in most cases. Moreover, the extrema‑based metric yields larger improvement factors. This behaviour is consistent with its definition: as a deliberately conservative, worst‑case estimator that relies on only two datapoints, the metric is disproportionately influenced by gains in frequency accuracy. Its strong sensitivity to intrinsic data-noise naturally results in a larger dataset‑to‑dataset variability, which is indeed what we observe (Appendix~\ref{app:detailed_table_improvement_factor}). While the standard‑deviation metric also shows some variability, it remains significantly more stable.

For this reason, we discarded the extrema‑based results when deriving our final averaged improvement factors, obtained by combining results from solar cycles 23 and 24. If individual frequencies are directly fitted, we find that magnetic activity effects are reduced by 24\%, 12\%, 30\%, and 38\% when moving respectively from 182.5 to 365 days, 365 to 730 days, 730 to 1460 days, and 365 to 1460 days. When fitting frequency separation ratios, the corresponding improvement factors are 13\%, 14\%, 20\%, and 31\%. From a physical standpoint, these numbers should be interpreted as indicative trends applicable to Sun-like observations with a similar main cycle duration. We also note that some modelling configurations that we tested displayed noticeable over‑ or under‑performance relative to these trends. Nevertheless, in the context of the PLATO mission, the key message remains clear: a continuous single-field 4‑year observing strategy would substantially smooth out magnetic activity effects for a distant solar analogue, the key target for the mission stellar accuracy requirements. Conversely, the currently favoured two-field 2+2‑year strategy provides only modest improvement, with little difference between 365‑day and 730‑day observing windows.

\section{Conclusions}
\label{sec:conclusions}
In this work, we examined how magnetic‑activity–induced biases evolve with observational time span by analysing 28 seismic‑age datasets constructed from BiSON and GOLF Sun‑as‑a‑star Doppler‑velocity observations, covering a broad range of observing baselines, inference methods, and mode selections. In Sect.~\ref{sec:imprint_MA_cycle}, we studied the imprint of the solar activity cycle on the inferred seismic ages through correlation analysis. In Sect.~\ref{sec:averaging_observing_baselines}, we assessed how effectively longer observing windows suppress these magnetic activity signatures.

Across all 28 configurations, a measurable trace of the solar cycle persists, even in ages inferred from the longest 1460‑day windows. With this baseline, the residual systematic uncertainty induced by magnetic activity remains at the $\sim$1--2.5\% level for fits of individual frequencies and $\sim$2.5--4.5\% for fits based on frequency separation ratios. Although these values lie well below PLATO's 10\% age‑accuracy requirement for a Sun‑like star, they are not negligible in the context of high‑precision asteroseismology and consequently underline the need for ongoing improvements in modelling approaches.

For fits to individual mode frequencies, including additional modes of low radial order weakens the correlation with magnetic activity in most cases, consistent with the 365‑day results of \citetalias{Betrisey2024_MA_Sun} and theoretical expectations since these modes are less affected by magnetic activity \citep[e.g.][]{Broomhall&Nakariakov2015}. This also leads to correspondingly smaller systematics (Appendix~\ref{app:evolution_MA_uncertainty}). Surface‑independent fits based on frequency separation ratios, however, display a more complex response: because ratios eliminate the need for an explicit surface‑effect prescription, magnetically induced perturbations propagate into age inferences in a non‑trivial manner. As a consequence, no straightforward uncertainty pattern emerges, in agreement with the conclusions of \citetalias{Betrisey2025_MA_Damping} with 365-day data.

The dependence on observing baseline reveals several regimes. On one hand, 365-day windows and 1460-day segments provide markedly stronger suppression of magnetic‑activity signatures than shorter baselines. This improvement arises from enhanced frequency determination (dominant at 365 days) and from increasingly efficient temporal averaging of the activity cycle (dominant at 1460 days). In contrast, 730‑day windows underperform relative to 365‑day ones in most cases: frequency accuracy has already plateaued, yet the observing window remains too short to smooth out cycle‑related variability. Combing results from solar cycles 23 and 24, the magnetic‑activity‑induced biases decrease by 24\%, 12\%, 30\%, and 38\% when transitioning respectively from 182.5 to 365 days, 365 to 730 days, 730 to 1460 days, and 365 to 1460 days for frequency-based fits. For fits employing frequency separation ratios, the corresponding improvements amount to 13\%, 14\%, 20\%, and 31\%.

In the context of the PLATO mission \citep{Rauer2025}, we therefore conclude that a continuous single-field four‑year observing programme would markedly mitigate magnetic activity effects for a distant solar analogue, the principal benchmark for the mission's stellar accuracy goals. Such a long baseline would also provide improved monitoring of surface activity through the photometric 
variability amplitude of the light curves \citep[see][]{Breton2024}, which may help constrain the correlation between this variability and the magnetic frequency shifts \citep[see][]{Santos2018}. In contrast, the currently favoured 2+2‑year configuration in two distinct fields provides only limited additional suppression, with minimal gains between the 365‑day and 730‑day observing segments. In practical terms, with a two‑year observing baseline, we expect activity biases at the $\sim$2--3\% level for frequency fits and $\sim$3.5--5.5\% for ratio fits, which are significant at the precision targeted by PLATO.

Altogether, our results underscore the importance of developing dedicated mitigation strategies for magnetic activity effects. Two complementary avenues appear promising. If independent constraints on cycle properties are available (e.g. $\rm S_{ph}$), through global activity indices, for example, which will become increasingly accessible thanks to coordinated follow‑up programs of space‑based surveys, the most direct approach is to pre‑correct the measured oscillation frequencies for cycle‑induced shifts. This could follow a similar methodology to \citet{Broomhall2009_correction} or build upon the theoretical framework developed by \citet{Dziembowski&Goode2004,Dziembowski&Goode2005}. However, at present, and for the early phases of PLATO, such prior knowledge will remain unavailable for the vast majority of targets. In this regime, an alternative is to design a surface‑effect-like prescription to absorb activity‑related perturbations in frequency‑ratio space, as proposed by \citetalias{Betrisey2025_MA_Damping}. It would also be worthwhile exploring non-solar regimes, as PLATO will observe broad range of solar‑type stars \citep{Goupil2024}, spanning a variety of magnetic‑cycle amplitudes, periods, and structural responses \citep[e.g.][]{Salabert2018}.

\begin{acknowledgements}
J.B. acknowledges funding from the SNF Postdoc.Mobility grant no. P500PT{\_}222217 (Impact of magnetic activity on the characterization of FGKM main-sequence host-stars). A.-M.B. has received support from STFC consolidated grant ST/X000915/1. H.D. and A.-M.B. acknowledge support from the Royal Astronomical Society for the Undergraduate Summer Research Bursary. S.N.B acknowledges support from PLATO ASI-INAF agreement no. 2022-28-HH.0 "PLATO Fase D". E.P. and R.A.G. acknowledges support from PLATO and GOLF CNES grants. O.K. acknowledges support by the Swedish Research Council (grant no. 2023-03667) and the Swedish National Space Agency. Finally, this work has benefited from financial support by CNES in the framework of its contribution to the PLATO mission.
\end{acknowledgements}

\bibliography{bibliography.bib}

\appendix

\section{Supplementary data on magnetic-activity systematic uncertainty}
\label{app:evolution_MA_uncertainty}
\begin{figure}[t]
\centering
\begin{subfigure}[b]{.37\textwidth}
  \includegraphics[width=.99\textwidth]{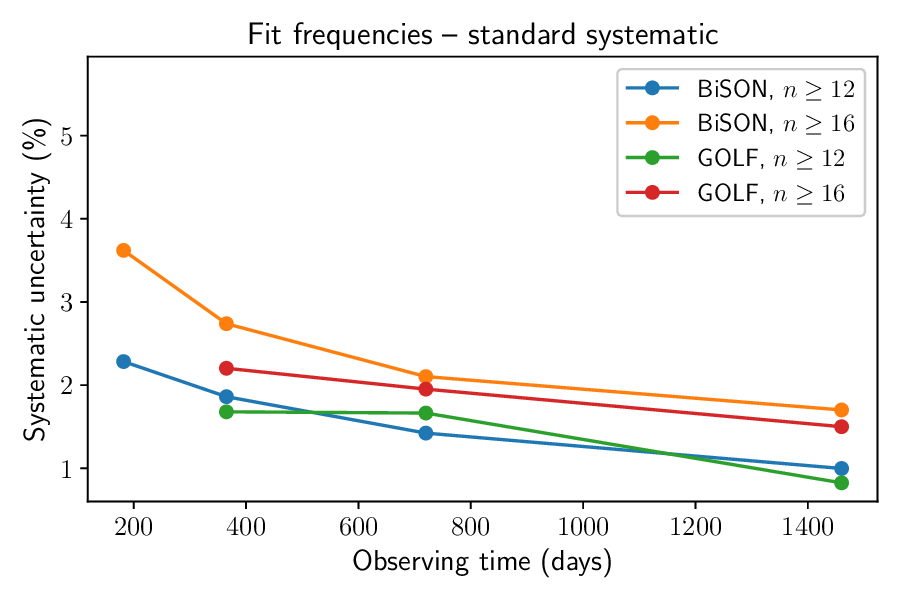}  
\end{subfigure}
\begin{subfigure}[b]{.37\textwidth}
  \includegraphics[width=.99\textwidth]{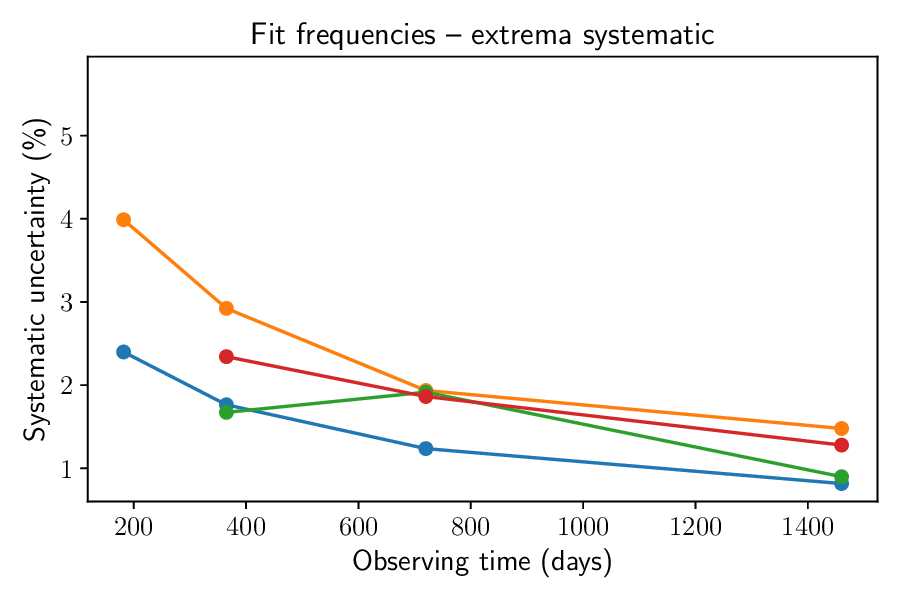}  
\end{subfigure}
\begin{subfigure}[b]{.37\textwidth}
  \includegraphics[width=.99\textwidth]{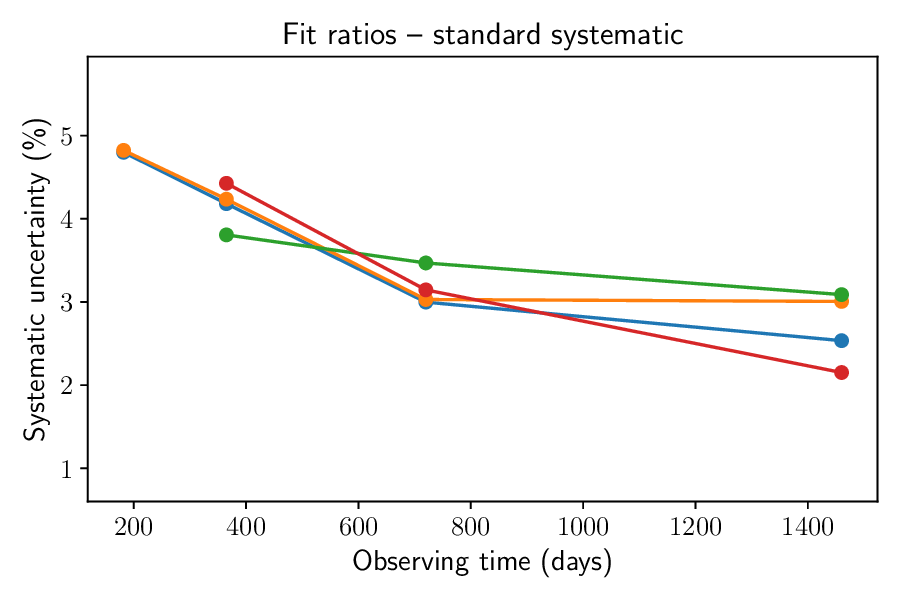}  
\end{subfigure}
\begin{subfigure}[b]{.37\textwidth}
  \includegraphics[width=.99\textwidth]{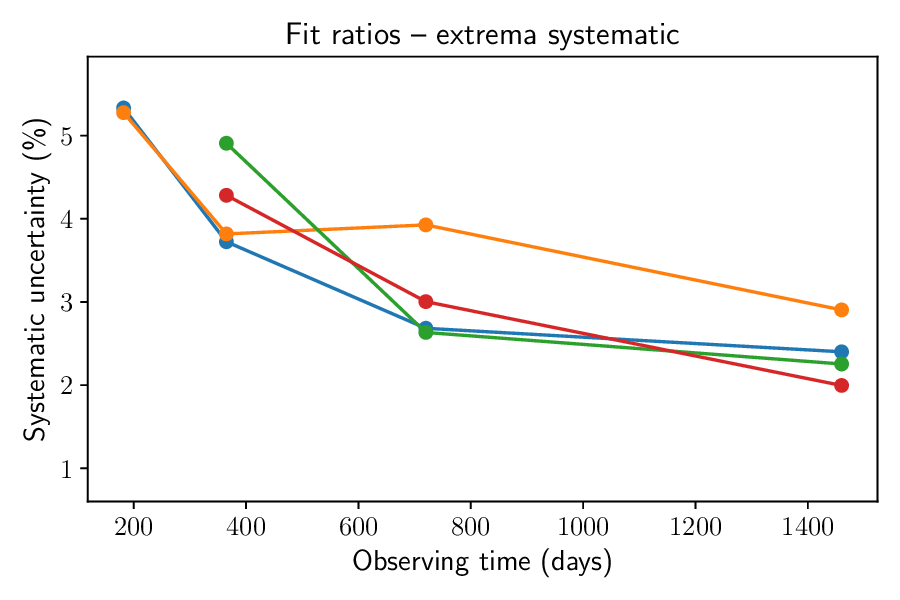}  
\end{subfigure}
\caption{Same as Fig.~\ref{fig:MA_systematic_cycle23} but for solar cycle 24.}
\label{fig:MA_systematic_cycle24}
\end{figure}

The systematic uncertainty associated with magnetic activity was quantified using two metrics from the literature: $\rm\sigma_{standard}$ \citepalias{Betrisey2025_MA_Inv,Betrisey2025_MA_Damping} and $\rm\sigma_{extrema}$ \citepalias{Betrisey2024_MA_Sun}. The evolution of the resulting systematic uncertainties over solar cycles~24 is shown in Fig.~\ref{fig:MA_systematic_cycle24}.

Because the metrics used in the main analysis are purely data-driven and can become unstable when only a limited number of datapoints are available, we also explored whether a parametric description of the magnetic-activity imprint could provide a more robust estimate of the systematic uncertainty. Such an approach would additionally offer a first step toward a surface-effect-like prescription capable of suppressing activity signatures in seismic inferences, particularly in cases where only a fraction of the cycle is observed or where the activity signal is intrinsically weak.

Because our datasets cover full solar activity cycles, we tested this idea using the Hathaway model \citep{Hathaway1994,Hathaway2015}, which provides a compact functional description of solar-cycle variability. We emphasise that
this experiment is exploratory and intended solely to assess feasibility, as PLATO's observing windows are in general too short to permit full-cycle coverage. The general form of the Hathaway function is given by
\begin{align}
\label{eq:hathaway}
H(t) = a\cdot \left(\frac{t-t_0}{b}\right)^3\left(e^{(t-t_0)^2/b^2}-c\right)^{-1}+d,
\end{align}
where $a$, $b$, and $c$ are free parameters related to the rise rate from minimum, the time between minimum and maximum, and the asymmetry of the cycle, respectively; $d$ is an offset, and $t_0$ marks the start of the cycle. The bias may then be estimated from the relative amplitude of the Hathaway function at activity maximum. The Hathaway model is optimally suited to high-quality activity proxies, such as the 10.7\,cm radio flux, which offer dense temporal sampling and high precision. Our seismic-age estimates, by contrast, exhibit lower precision and provide comparatively sparse sampling of the activity cycle, particularly for the 730-day and 1460-day segments. Moreover, although the seismic ages correlate strongly with the activity proxies, the correspondence is not fully linear. As a result, fitting the full Hathaway model, including its asymmetry term, proved unfeasible for our datasets.

To assess whether a simplified version might still be informative, we repeated the fit with the asymmetry parameter fixed to $c = 0$. In that case, the relative amplitude can be found analytically. Setting $x = (t - t_0)/b$ and solving
\begin{align}
\frac{d}{dx}H(x) = 0 \Rightarrow x = \pm \sqrt{\frac{3}{2}},
\end{align}
the positive root gives the location of the maximum, and the corresponding relative amplitude is
\begin{align}
\left(\frac{3}{2}\right)^{3/2}e^{-3/2}\cdot a \simeq 0.409\,a.
\end{align}
While this simplified model converged, its performance was poor, especially for Cycle~24. This outcome is expected: Cycle~24 exhibits lower amplitude and appears noisier than Cycle~23, exacerbating the limitations already imposed by the data quality. Even for Cycle~23, where the results initially appeared reasonable and produced bias estimates broadly consistent with the other metrics, the solution remained overly sensitive to the offset parameter $d$. This sensitivity prevents the simplified Hathaway-based estimate from being used in a robust quantitative manner.

\section{Detailed table for improvement factor}
\label{app:detailed_table_improvement_factor}
Table~\ref{tab:improvement_factor_full} summarises the improvement factors obtained for all configurations explored in this study. These configurations span different instruments (GOLF and BiSON), observing baselines (182.5, 365, 730, and 1460 days), inference strategies (a surface-dependent approach based on a direct fit to individual frequencies, and a surface-independent approach based on frequency separation ratios), and mode selections ($n \geq 12$ and $n \geq 16$). In total, this parameter space yields 28 distinct seismic-age datasets.

\begin{table*}[h!]
\centering
\caption{Detailed table for the improvement factor due to an increase of the observing time for solar cycles 23 and 24.}
\begin{tabular}{lcccccccc}
\hline \hline 
& \multicolumn{4}{c}{Fit frequencies} & \multicolumn{4}{c}{Fit ratios} \\
 \cmidrule[0.4pt](lr){2-5} \cmidrule[0.4pt](l){6-9}
 & \multicolumn{2}{c}{Cycle 23} & \multicolumn{2}{c}{Cycle 24} & \multicolumn{2}{c}{Cycle 23} & \multicolumn{2}{c}{Cycle 24} \\
  \cmidrule[0.4pt](lr){2-3} \cmidrule[0.4pt](lr){4-5} \cmidrule[0.4pt](lr){6-7} \cmidrule[0.4pt](l){8-9}
 & $\mathcal{A}_{standard}$ & $\mathcal{A}_{extrema}$ & $\mathcal{A}_{standard}$ & $\mathcal{A}_{extrema}$ & $\mathcal{A}_{standard}$ & $\mathcal{A}_{extrema}$ & $\mathcal{A}_{standard}$ & $\mathcal{A}_{extrema}$ \\
\hline 
$182.5 \rightarrow 365$ days &  &  &  &  &  &  &  &  \\ 
BiSON, $n \geq 12$ & 31.9\% & 32.3\% & 18.5\% & 26.5\% & 8.8\% & 4.8\% & 12.9\% & 30.2\% \\ 
BiSON, $n \geq 16$ & 21.1\% & 32.4\% & 24.3\% & 26.7\% & 18.9\% & 24.8\% & 12.2\% & 27.6\% \\ 
GOLF, $n \geq 12$ & - & - & - & - & - & - & - & - \\ 
GOLF, $n \geq 16$ & - & - & - & - & - & - & - & - \\ 
Average & 26.5\% & 32.3\% & 21.4\% & 26.6\% & 13.9\% & 14.8\% & 12.5\% & 28.9\% \\ 
\hline 
$365 \rightarrow 730$ days &  &  &  &  &  &  &  &  \\
BiSON, $n \geq 12$ & 12.0\% & 47.0\% & 23.5\% & 29.9\% & 13.5\% & 18.2\% & 28.3\% & 27.9\% \\ 
BiSON, $n \geq 16$ & 7.8\% & 13.3\% & 23.2\% & 33.8\% & 6.0\% & 19.9\% & 28.5\% & -2.8\% \\ 
GOLF, $n \geq 12$ & 16.8\% & 21.6\% & 0.9\% & -14.5\% & -4.8\% & 2.2\% & 8.9\% & 46.3\% \\ 
GOLF, $n \geq 16$ & 1.0\% & 18.5\% & 11.4\% & 20.5\% & 3.1\% & 11.7\% & 29.0\% & 29.9\% \\ 
Average & 9.4\% & 25.1\% & 14.8\% & 17.4\% & 4.5\% & 13.0\% & 23.7\% & 25.3\% \\
\hline 
$730 \rightarrow 1460$ days &  &  &  &  &  &  &  &  \\  
BiSON, $n \geq 12$ & 22.0\% & 24.8\% & 29.9\% & 34.0\% & 27.8\% & 44.4\% & 15.5\% & 10.5\% \\ 
BiSON, $n \geq 16$ & 19.9\% & 30.7\% & 19.1\% & 23.6\% & 27.9\% & 36.7\% & 0.8\% & 26.0\% \\ 
GOLF, $n \geq 12$ & 48.1\% & 57.9\% & 50.4\% & 53.1\% & 22.4\% & 38.8\% & 11.0\% & 14.4\% \\ 
GOLF, $n \geq 16$ & 24.9\% & 37.6\% & 23.2\% & 31.3\% & 20.9\% & 24.4\% & 31.6\% & 33.6\% \\ 
Average & 28.7\% & 37.8\% & 30.7\% & 35.5\% & 24.8\% & 36.1\% & 14.7\% & 21.1\% \\ 
\hline 
$365 \rightarrow 1460$ days &  &  &  &  &  &  &  &  \\ 
BiSON, $n \geq 12$ & 31.3\% & 60.1\% & 46.4\% & 53.7\% & 37.6\% & 54.5\% & 39.4\% & 35.5\% \\ 
BiSON, $n \geq 16$ & 26.2\% & 40.0\% & 37.9\% & 49.4\% & 32.3\% & 49.3\% & 29.0\% & 23.9\% \\ 
GOLF, $n \geq 12$ & 56.8\% & 67.0\% & 50.9\% & 46.3\% & 18.7\% & 40.2\% & 18.9\% & 54.1\% \\ 
GOLF, $n \geq 16$ & 25.7\% & 49.2\% & 31.9\% & 45.4\% & 23.4\% & 33.2\% & 51.4\% & 53.4\% \\ 
Average & 35.0\% & 54.1\% & 41.8\% & 48.7\% & 28.0\% & 44.3\% & 34.7\% & 41.7\% \\ 
\hline 
\end{tabular}
\label{tab:improvement_factor_full}
{\par\small\justify\textbf{Notes.}  The improvement factor is defined as $\mathcal{A}=1-\sigma_a/\sigma_b$, where $\sigma$ is the systematic uncertainty associated to magnetic activity and $a>b$ are two observing times. Two methods were used to compute $\sigma$: i) the definition from \citetalias{Betrisey2025_MA_Inv} and \citetalias{Betrisey2025_MA_Damping} based on the standard deviation of the dataset, and ii) the definition from \citetalias{Betrisey2024_MA_Sun} based on the dataset extrema. \par}
\end{table*}

\end{document}